\documentclass[preprint,preprintnumbers,showpacs,amssymb,amsmath,secnumarabic,tightenlines,nofootinbib,12pt]{revtex4}
\usepackage{bm}
\usepackage{psfig}
\def\pdf{{\it pdf}}
\def\cdf{{\it cdf}}
\def\babar{\mbox{\sl B\hspace{-0.4em} {\scriptsize\sl A}\hspace{-0.4em} B\hspace{-0.4em} {\scriptsize\sl A\hspace{-0.1em}R}}}
\def\lkh{\mbox{$\mathcal L$}}
\def\kolm{Kolmogorov-Smirnov}

\long\def\simplex#1#2#3#4{
\begin{figure}[#1]
   \begin{center}
   \quad\\[-2.5cm]
   \quad\vskip  2.0 cm
   \hbox{
   \quad 
   \parbox[t]{6.5cm}{ \psfig{file=#2,width=8.5cm}
   \caption[c]{\small  \label{fig:#3} #4 } }
   }
   \quad
   \end{center} 
\end{figure}
}
\long\def\duplex#1#2#3#4#5{
\begin{figure}[#1]
   \begin{center}
   \vskip +1cm
   \quad\\[-1.0cm]
   \quad
   \hbox{\hskip -0cm
   \quad 
   \parbox[t]{5.5cm}{ \psfig{figure=#2,width=6.5cm}
   } 
   \quad
   \parbox[t]{5.5cm}{ 
   \psfig{figure=#3,width=6.5cm} }
   }
   \caption[c]{\small \label{fig:#4} #5 }
   \quad
   \end{center} 
\end{figure}
}

\begin{document}

\preprint{CALT-HEP-68-2443}

\title{Estimation of Goodness-of-Fit in Multidimensional Analysis 
	Using Distance to Nearest Neighbor
\footnote{Work partially supported by Department of Energy under Grant
DE-FG03-92-ER40701.}
}
\author{Ilya Narsky}
\email{narsky@hep.caltech.edu}
\affiliation{California Institute of Technology}
\date{\today}

\begin{abstract}
A new method for calculation of goodness of multidimensional fits in
particle physics experiments is proposed. This method finds the
smallest and largest clusters of nearest neighbors for observed data
points. The cluster size is used to estimate the goodness-of-fit and
the cluster location provides clues about possible problems with data
modeling. The performance of the new method is compared to that of the
likelihood method and \kolm\ test using toy Monte Carlo studies. The
new method is applied to estimate the goodness-of-fit in a $B\to Kll$
analysis at \babar.
\end{abstract}

\pacs{02.50.-r, 02.50.Ng, 02.50.Sk.}

\maketitle

\section{Introduction}

Fits are broadly used in analysis of particle physics experiments.  If
sufficient statistics is accumulated, one usually plots observed data
as a histogram and overlays an expected histogram or modeling
function. The goodness-of-fit is then estimated by taking the
deviation of the observed number of events in each bin from the
expected number of events in the same bin, summing the squares of
these deviations to form a $\chi^2$ value and hence compute a
significance level.  This procedure assumes that the bin contents are
normally distributed, which is true only asymptotically in the large
statistics limit.  In low-statistics experiments, one typically
observes zero or just a few events per bin, and this procedure does
not produce a reliable result. One must then use other methods.
Unbinned maximum likelihood methods, discussed below, have been
recently used in such situations in \babar\ and elsewhere.  We also
discuss \kolm\ test, another well-known method.

In this note, we concentrate on the most general problem: 
how to test a distribution
in question against every reasonable alternative hypothesis. In other
words, the null and alternative hypotheses are stated as follows:\\
\begin{tabular}{rl}
$H_0:$ & the observed data obey the expected distribution.\\
$H_1:$ & the observed data obey some other unknown but plausible 
	distribution.\\
\end{tabular}\\
The goodness-of-fit, $1-\alpha$, is defined as the confidence level of
the null hypothesis, and $\alpha$ is therefore a Type~I error.  We
remind the reader that the Type~II error is traditionally defined as
the probablility of accepting the null hypothesis if the alternative
hypothesis is true. 

This definition of the problem conforms with the standard $\chi^2$
test of binned data. Indeed, the $\chi^2$ test computes discrepancy
between expected and observed probability density functions (\pdf's)
without imposing constraints on the alternative hypothesis.  We do not
discuss examples, where the alternative hypothesis $H_1$ can be stated
in a more specific form, e.g., testing normality versus
uniformity. Our goal is to propose a new generic procedure applicable
to unbinned fits.

It is not possible to design a versatile procedure applicable to all
problems. For example, we can always choose the alternative
distribution to be a set of $\delta$-functions positioned precisely at
the observed experimental points. In this case, the null hypothesis is
inferior to the alternative hypothesis and the null hypothesis is
rejected. This simply reflects the fact that the Type~II error is
undefined for the generic test stated in the previous paragraph.
We would like to keep our procedure as generic as possible. Yet if
more information about the alternative hypothesis is available, it
should be possible to design a more powerful test for this specific
alternative. 

We note that the standard $\chi^2$ binned test computes an average
deviation of observed data from the expected density. However, in many
experiments it is useful to focus on the maximal deviation instead of
the average one.  Consider, for example, fitting a one-dimensional
histogram divided into 20 bins in the range $[-10,10]$ to the sum of a
standard normal \pdf\ $N(0,1)$ with zero mean and unit variance and a
uniform \pdf, as shown in Fig.~\ref{fig:3chisqs}. The normal \pdf\
represents signal (for example, mass of a certain resonance) and the
uniform \pdf\ represents background, with the magnitude of each
component fixed to 100 entries. The $\chi^2$ deviation, computed as
$\sum_{bins}(N_{expected}-N_{observed})^2/N_{expected}$, is 19.34 per
20 degrees of freedom for each of the three fits, which results in a
goodness-of-fit value of 50\%. Hence, the procedure treats all fits as
those of equal quality.
In reality, of course, the experimenter will treat each fit in a
different way. The top fit will be likely considered as ``good''. The
middle fit will likely raise concern about a large background
fluctuation in one bin. 
The bottom fit will likely make the experimenter suspect that
the signal is not well modeled by the normal standard \pdf\ with an
area of 100. In fact, the experimenter is not really concerned about the
$\chi^2$ deviation averaged over all bins. The more interesting
question is: what are the bins that give largest $\chi^2$ deviations
from expected values and how probable are these deviations? The method
proposed in this note is designed to answer both these questions for
unbinned fits. 

\simplex{htbp}{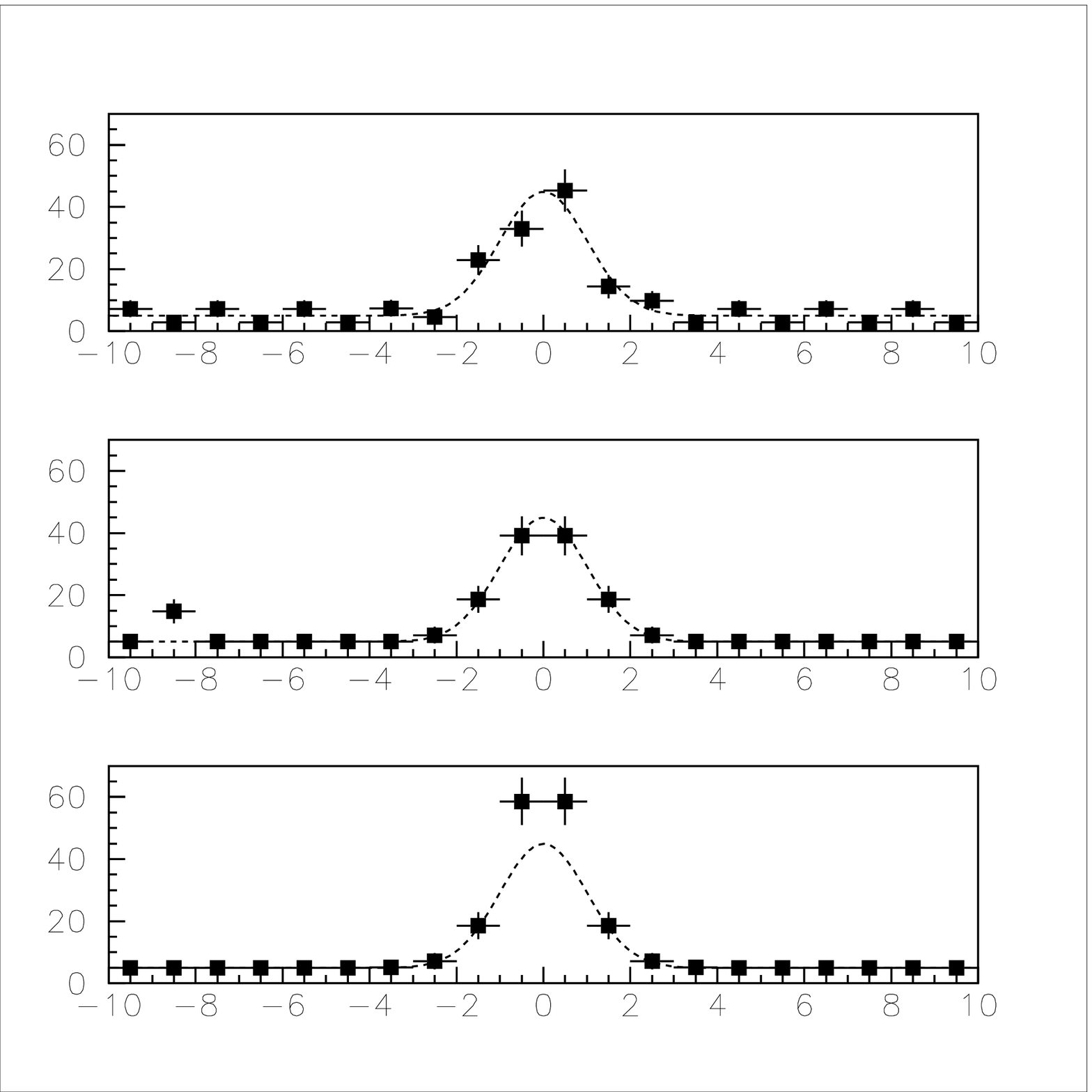}{3chisqs}{ Three fits with 50\%
goodness-of-fit values computed using the standard $\chi^2$
method for binned data. Top plot --- the $\chi^2$ deviations are
distributed uniformly over the bins; middle plot --- the $\chi^2$
deviation is entirely due to one bin at the left edge of the
histogram; bottom plot --- the $\chi^2$ deviation is produced by the
two central bins.}

\section{Maximum Likelihood Value Test}
\label{sec:mlv}

The Maximum Likelihood Value (MLV) test is laid out in the \babar\
Statistics Report~\cite{bbr_rep_gof}. For any quantity $x$ that
characterizes fit quality, the goodness-of-fit is given by
\begin{equation}
\label{eq:mlvdef}
1-\alpha = 1 - \int_{f(x|H_0)>f(x_{obs}|H_0)} f(x|H_0) dx\ ,
\end{equation}
where 
$x_{obs}$ is the value of $x$ observed in the fit to the data, and $f(x|H_0)$
is the \pdf\ of quantity $x$ under the null hypothesis. For the MLV
test, the quantity of interest is the likelihood and so $\lkh$
replaces $x$ in the equation above.

By construction, the MLV test can be only used to discriminate against
a specific class of alternative hypotheses. Data are fitted to the
density $f(x|\theta)$ and an estimate of the parameter
$\theta=\theta_0$ is obtained from the fit. Then the null hypothesis
$H_0:\theta=\theta_0$ is tested against the alternative hypothesis
$H_1:\theta\neq\theta_0$.  Note that the overall validity of the
density $f(x|\theta)$ is never questioned.  If the data are drawn from
a drastically different \pdf, this test can produce a meaningless
result.

Consider, for instance, fitting a one-dimensional random sample to a
standard normal \pdf\ $N(0,1)$. In reality, however, the data are
drawn from a sum of two narrow normal \pdf's placed two units apart:
$N(-1,0.01)$ and $N(+1,0.01)$, as shown in
Fig.~\ref{fig:g_vs_2g}. Distributions of likelihood values computed
under the null hypothesis for events drawn from the standard normal
\pdf\ $N(0,1)$ and events drawn from the sum of two normal \pdf's are
shown in Fig.~\ref{fig:g_vs_2g}. Likelihood values computed under the
null hypothesis $N(0,1)$ for the sum of two narrow normal \pdf's are
always consistent with the null hypothesis. The procedure does not
have any discriminative power and the obtained fit always produces a
reasonable goodness-of-fit value, even though the null hypothesis is
clearly wrong. In this example an experimenter can easily find the
problem by visual comparison of the distributions, but in the real
world of multidimensional distributions such a comparison would be
harder to make.

\duplex{htbp}{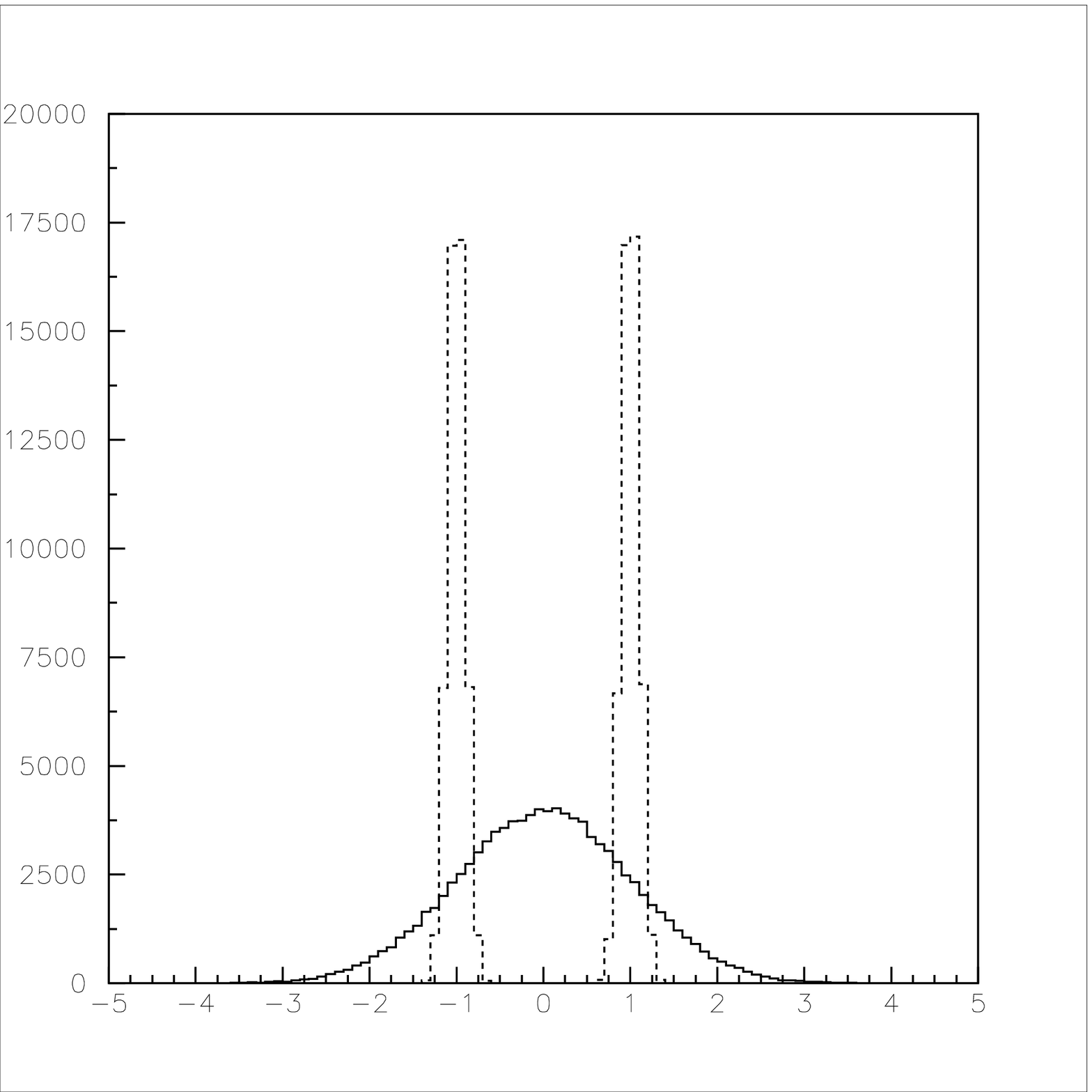}{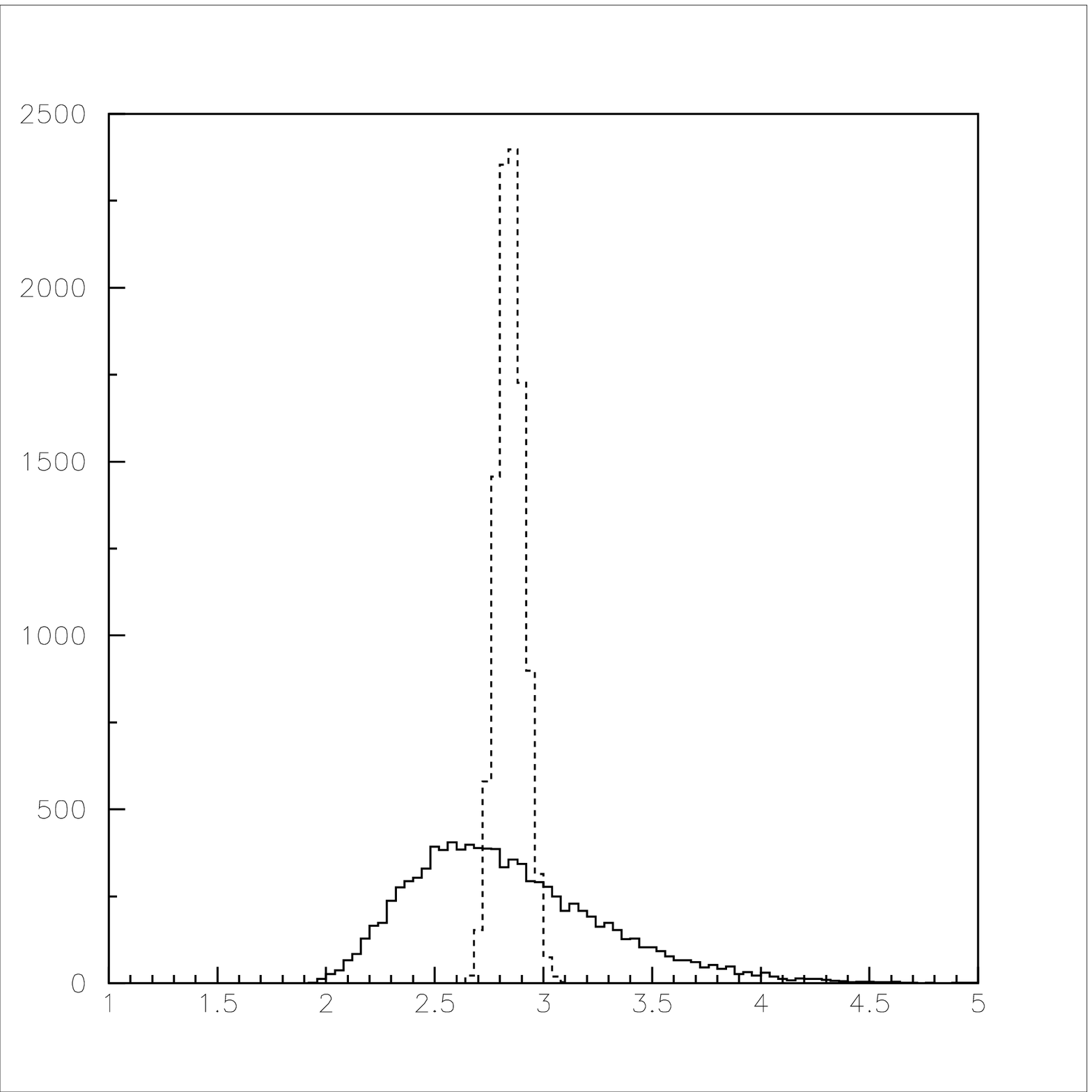}{g_vs_2g}{
Densities for a standard normal \pdf\ $N(0,1)$ (solid line) and
a sum of two narrow normal \pdf's $N(-1,0.01)$ and $N(+1,0.01)$
(dashed line) are shown on the left.
$-2\log\lkh_0$ distributions computed under the null hypothesis
$N(0,1)$ for both \pdf's are shown on the right for 10,000
toy MC experiments with 10 events in each experiment.}

Why did the MLV test fail to reject the null hypothesis for the
random sample described in the previous paragraph?  
Because the alternative hypothesis $H_1$ was not ``every other
plausible distribution'' but ``another normal distribution''.  The
price for this assumption was a futile test.  It is true that the
procedure would also discriminate against certain non-normal
distributions. But it would work by accident, not by design. 

Another good example is a test of uniformity. 
Maximum likelihood methods are useless here because under uniformity
the likelihood value is constant, no matter how points are
distributed.\footnote{Unless there is an experimental point observed
outside the range of definition of the uniform \pdf.}

\section{Generic Tests}

\subsection{Outline}

In the previous section, we established that the likelihood method
does not address the problem stated in the Introduction. A more
versatile approach is to test the null hypothesis without making
specific assumptions about its alternative. We refer to such tests as
``generic''. 

In Section~\ref{sec:statlit}, we briefly discuss information about
generic tests that can be found in the statistics literature. Then we
proceed with discussion of the \kolm\ test, a well-known generic
approach, in Section~\ref{sec:kolm} and propose a new method in
Section~\ref{sec:dtnn}. We emphasize that any generic test can be
standardized by transforming the density of interest to uniform and
performing a test of uniformity. The transformation to a uniform
density is described in Section~\ref{sec:trans} and the definition of
uniformity is discussed in Section~\ref{sec:unif}. The transformation
to a uniform density is not required for the \kolm\ test but is
essential for the proposed method. Subtleties related to the
non-uniqueness of the uniformity transformation are discussed in
Section~\ref{sec:invariance}.

\subsection{Statistics Literature}
\label{sec:statlit}

There is a great amount of statistics literature on goodness-of-fit
tests. 
Unfortunately,
a great fraction of this literature is useless to us because of one of
the following reasons:
\begin{itemize}
	\item The discussed problem is too specific, e.g., testing
	one specific type of \pdf\ against another specific type of
	\pdf.
	\item Asympotic approximations, e.g., the central limit theorem, are
	used. 
	\item Authors concentrate on designing an analytic tool
	(inevitably based on some approximation) and dismiss MC
	simulation. 
\end{itemize}
We, on the other hand, would like to have a generic approach for
unbinned fits with 
small numbers of events. We can rely on MC generators; hence,
analyticity of the solution is not an issue.

A well-known method that complies with these requirements is the
\kolm\ test. However, the \kolm\ test lacks sensitivity for a broad
class of alternative hypotheses.

To our knowledge, the distance-to-nearest-neighbor method proposed in
Section~\ref{sec:dtnn} has not been described anywhere in the
literature. As the idea seems obvious, it is quite possible that we
are simply reinventing the wheel. But we hope that this wheel is worth
reinventing.

\subsection{Transformation to Uniform Density}
\label{sec:trans}

Before we proceed with discussion of various methods, we note that the
problem can be standardized by transforming any $n$-dimensional \pdf\
in question to uniform. This transformation offers a number of
advantages:
\begin{itemize}
	\item All problems are described by the same formalism.
	\item The practical task of generating toy MC experiments
	is easily solved with a uniform random number generator.
	\item It is easy to implement this transformation numerically
	in the absence of an analytic model for the \pdf.
\end{itemize}

Any $n$-dimensional random vector $(x^{(1)},x^{(2)},...,x^{(n)})$ with
a joint \pdf\ $f(x^{(1)},x^{(2)},...,x^{(n)})$ can be transformed to a
vector $(u^{(1)},u^{(2)},...,u^{(n)})$ uniformly distributed on an
$n$-dimensional unit cube $0\leq u^{(i)}\leq 1;\ i=1,2,...,n$.  This
transformation is given by
\begin{eqnarray}
\label{eq:tounif}
\left\{
\begin{array}{lcl}
u^{(1)} & = & \int_{-\infty}^{x^{(1)}} f_1(t,x^{(2)},x^{(3)},...,x^{(n)}) dt / 
	      f_2(x^{(2)},x^{(3)},...,x^{(n)}) \\
u^{(2)} & = & \int_{-\infty}^{x^{(2)}} f_2(t,x^{(3)},x^{(4)},...,x^{(n)}) dt / 
	      f_3(x^{(3)},x^{(4)},...,x^{(n)}) \\
        &   & ... \\
u^{(n-1)} & = & \int_{-\infty}^{x^{(n-1)}} f_{n-1}(t,x^{(n)}) dt / 
	        f_n(x^{(n)}) \\
u^{(n)} & = & \int_{-\infty}^{x^{(n)}} f_n(t) dt
\end{array}
\right.
\end{eqnarray}
where
\begin{eqnarray}
\left\{
\begin{array}{lcl}
f_1(x^{(1)},x^{(2)},...,x^{(n)}) & = & f(x^{(1)},x^{(2)},...,x^{(n)}) \\
f_2(x^{(2)},x^{(3)},...,x^{(n)}) & = & \int_{-\infty}^{+\infty} 
			f_1(x^{(1)},x^{(2)},...,x^{(n)}) dx^{(1)} \\
f_3(x^{(3)},x^{(4)},...,x^{(n)}) & = & \int_{-\infty}^{+\infty} 
			f_2(x^{(2)},x^{(3)},...,x^{(n)}) dx^{(2)} \\
                     &   & ... \\
f_n(x^{(n)}) & = & \int_{-\infty}^{+\infty} 
			f_{n-1}(x^{(n-1)},x^{(n)}) dx^{(n-1)}
\end{array}
\right.
\end{eqnarray}

This transformation is one-to-one for a strictly positive \pdf\
$f(\vec{x})$.

The cumulative density function (\cdf) for an $n$-dimensional uniform
distribution is simply
\begin{equation}
F(\vec{u}) = \prod_{i=1}^{n} u^{(i)}\ .
\end{equation}

\subsection{\kolm\ Test}
\label{sec:kolm}

A generic method broadly known to physicists is the \kolm\ test. The
\kolm\ statistic for a random sample $\vec{x}_1,\vec{x}_2,...,\vec{x}_N$ 
of $n$-dimensional vectors
$\vec{x}=(x^{(1)},x^{(2)},...,x^{(n)})$ with a \cdf\ $F(\vec{x})$ is
given by~\cite{saunders,justel}
\begin{equation}
\label{eq:kolm}
K_N(F) = \mbox{sup}_{\vec{x} \in V_n} |F(\vec{x})-F_{obs}(\vec{x})|\ ,
\end{equation}
where $V_n$ is an $n$-dimensional domain for the \cdf\ $F(\vec{x})$,
and $F_{obs}(\vec{x})$ is the experimentally observed \cdf.  
The null
hypothesis is accepted if $K_N(F_0)$ is small and rejected if
$K_N(F_0)$ is large, where $F_0$ is a \cdf\ for the null hypothesis.


Because the \kolm\ test compares cumulative densities, it lacks sensitivity
to fluctuations within small clusters. Consider, for example, two sets
of points on a unit interval $0\leq x\leq 1$:\\
\begin{tabular}{ll}
Set 1: & $x_1=1/4,\ x_2=1/2,\ x_3=3/4,\ x_4=1$ \\
Set 2: & $x_1=x_2=1/4,\ x_3=x_4=3/4$           \\
\end{tabular}\\
Which one of these sets looks more uniform? The \kolm\ test cannot
differentiate between these two because the statistic~(\ref{eq:kolm}) is
1/4 under uniformity in both cases.

\subsection{What Is ``Uniform''?}
\label{sec:unif}

In fact, the question we asked in the previous section is not so
simple. Can we indeed make a statement about which set is more likely
to be drawn from a uniform distribution? The answer is: it depends.

Suppose we search for a heavily-suppressed decay using \babar\
data. We plot the mass distribution and we are convinced that
background in our analysis is flat. We would like to know if there is
an indication of any mass peaks in the plotted data. The data points
in Set~1 from the previous section are equally spaced while the data
points from Set~2 are grouped together in two clusters. Hence, Set~1
looks more uniform than Set~2.

Consider now another example. We have a detector that registers
ionizing particles. We would like to test the randomness of the
particle flux, that is, the exponentiality of the distribution of time
intervals between consecutive events. However, after an event is
registered, the detector becomes inactive for a certain period of
time. If the expected time interval between two consecutive events is
much less than the detector's deadtime, the device will trigger at
fixed time intervals. This would indicate that the process is not
exponential but periodic. On these grounds, we would conclude that
Set~1 looks less uniform than Set~2.

We obtained two opposite answers to the same question. Of course, the
question was not the same; in effect, these were two different
questions. In the first example, the vaguely stated alternative
hypothesis was ``presence of peaks in the data''. In the second
example, it was ``equidistant points on a finite interval''.  We
cannot design a test that gives the right answer for every possible
problem. Nevertheless, it would be good to have a procedure which is
more sensitive to clustering of data than the \kolm\ test is.

\subsection{Distance-to-Nearest-Neighbor Test of Uniformity}
\label{sec:dtnn}

The idea of using the distance to nearest neighbor for a test of
uniformity is not new~\cite{clark_evans,ripley,cuzick}.  For each data
point, $\vec{u}_i=(u_i^{(1)},u_i^{(2)},...,u_i^{(n)})$, in an
$n$-dimensional unit cube we find the nearest neighbor, $\vec{u}_j$,
and compute the distance, $d_{ij}=|\vec{u}_i-\vec{u}_j|$.  Uniformity
is tested by comparing observed values of $d_{ij}$ with those expected
for a uniform distribution.  In a more general approach, one can use
an average distance $d_{i}^{(m)}$ to $m>1$ nearest neighbors. In
Refs.~\cite{clark_evans,ripley,cuzick}, discussion revolves around
using moments of distributions of distances $d_{i}^{(m)}$ as test
statistics.  We propose a test of uniformity based on minimal and
maximal values of the distance $d_{i}^{(m)}$ to $m$ nearest
neighbors. It is intuitively clear that such test should be more
sensitive to maximal deviations of observed data from the tested \pdf\
than the \kolm\ test is.

A similar approach would be to use maximal and minimal volumes of
Voronoi regions. A Voronoi region for a given observed point
$\vec{u}_i$ is defined as a set of points inside the $n$-dimensional
unit cube which are closer to $\vec{u}_i$ than to any other observed
point $\vec{u}_j,\ j\neq i$. Voronoi regions have been used by the
Sleuth algorithm~\cite{sleuth} to search for new physics at the D0
experiment.  In essence, Sleuth computes the probability of observing
one data point in each Voronoi region based on the expectation value
for background and marks Voronoi cells with low probabilities as
candidates for a new physics signal. This method addresses the same
question: how consistent are observed data with a null hypothesis,
where the null hypothesis is defined as ``background events only''. To
a zeroth order, the volume of a Voronoi region around $m+1$ points is
proportional to the average size $d_{i}^{(m)}$ of the
cluster. Therefore, both methods for goodness-of-fit estimation are
expected to produce similar results. This is confirmed by MC tests
described in Section~\ref{sec:tests}.  However, using distance to
nearest neighbor is computationally simpler because the construction
of Voronoi regions can be avoided.

\subsection{Invariance of Test Statistic under Uniformity Transformation}
\label{sec:invariance}

The transformation to uniformity is not unique, even if we limit the
problem to continuous mappings. A transformation $\vec{x}\to\vec{u}$
is continuous if two infinitely close points are mapped onto two
infinitely close points, i.e., $\lim_{|\vec{x}_i-\vec{x}_j| \to 0}
|\vec{u}_i-\vec{u}_j| = 0$.  Consider, for example, a uniform \pdf\ on
a unit circle $f(r,\phi)=1/\pi;\ 0\leq r\leq 1,\ 0\leq \phi \leq
2\pi$. The joint \pdf\ of random variables
\begin{equation}
\begin{array}{lcll}
r'    & = & r                & \\
\phi' & = & \phi + \alpha r; & 0<|\alpha|<\infty
\end{array}
\end{equation}
is also uniform on the unit circle. However, this transformation does
not conserve distance between two points. Another example is
relabeling of variables $x^{(i)}$ in transformation~(\ref{eq:tounif})
for a non-factorizable \pdf\ in $n>1$ dimensions. 

It is clear that all
possible transformations to uniformity do not necessarily produce
identical values either for the \kolm\ statistic or the distance to
nearest neighbor. Inevitably, the value of goodness-of-fit for a specific
set of experimental data depends on the choice of transformation.
We do not consider this circumstance as a major obstacle. In many
problems, it is possible to find a reasonable transformation to
uniformity that preserves the natural metric of the experiment.

In many particle physics experiments, observation variables are
independent or weakly correlated. The \pdf\ of interest is therefore
factorizable or close to such. In this case,
transformation~(\ref{eq:tounif}) is reduced to
$u^{(i)}=F_i(x^{(i)}),\ i=1,2,...,n$, where $F_i$ is a
marginal \cdf\ for $i$th component. The transformation above is the
most obvious and natural choice.  In other experiments, the \pdf\ can
be transformed to a factorizable one. For example, a two-dimensional
normal \pdf\ can be rotated to align the axes of the normal elliptic
contour with the coordinate axes.

If the \pdf\ is severely non-factorizable, one can split $n$
observation variables into $k$ mutually independent (or weakly
correlated) groups with $n_i,\ i=1,2,...,k,$ variables in each group,
$n=n_1+n_2+...+n_k$.  Within each group, variables are strongly
correlated and the marginal $n_i$-dimensional \pdf\ cannot be
factorized. To obtain a test statistic invariant under relabeling of
observation variables $x^{(i)}$ in transformation~(\ref{eq:tounif}),
one would have to try all $n_i!$ permutations of variables within each
group.  For example, the minimal distance to nearest neighbor would be
chosen as the minimum of all distances to nearest neighbor in these
$n_i!$ permutations. This method was proposed~\cite{justel} for a
multidimensional \kolm\ test. We simply restate it here in reference
to the distance-to-nearest-neighbor approach.

\section{Tests}
\label{sec:tests}

We consider four two-dimensional \pdf's $f(x,y)$:
\begin{itemize}
	\item normal \pdf\ 
	$N(\mu_X=0,\mu_Y=0,\sigma_X^2=1,\sigma_Y^2=1,\rho=0)$ 
	with zero means, unit variances and 
	zero correlation between $x$ and $y$ 
	\item narrow normal \pdf\ $N(0,0,0.25,0.25,0)$
	\item sum of two narrow normal \pdf's $N(-1.3,0,0.01,0.01,0)$ 
	and $N(+1.3,0,0.01,0.01,0)$
	\item uniform \pdf\ defined on a square 
	$-5\leq x\leq 5;\ -5\leq y\leq 5$
\end{itemize}
For each density, we run 10,000 toy MC experiments with 10 events per
experiment. We use the standard normal \pdf\ $N(0,0,1,1,0)$ as the
null hypothesis (except one example, as discussed below) and plot in
Fig.~\ref{fig:4lkh} likelihood values $-2\log\lkh_0$ computed under
the null hypothesis for all \pdf's. Assuming the null hypothesis, we
apply uniformity transformation to each MC experiment and plot values
of the \kolm\ statistic for all \pdf's in Fig.~\ref{fig:4kolm}.  We
also plot two-dimensional distributions of maximal versus minimal
distance to nearest neighbor in Fig.~\ref{fig:4dist}. We use these MC
distributions to estimate Type~II errors for hypothesis tests at a
given confidence level against each alternative to the null
hypothesis. The confidence levels and errors are shown in Table~1. We
repeat this exercise treating the uniform \pdf\ as the null hypothesis
and testing it against the standard normal \pdf\ $N(0,0,1,1,0)$. This
result is also shown in Table~1 and Fig.~\ref{fig:udist}.

With the definitions of the Type~II error and confidence level shown
in the Introduction,
the smaller is the Type~II error for a fixed confidence
level, the more powerful is the test.

We compared results obtained by the distance-to-nearest-neighbor method
to those obtained through Voronoi regions and found no significant
difference. 

The maximum likelihood method is very efficient for discriminating one
normal \pdf\ against another and against a uniform \pdf\ which can be
considered as a limiting case of a normal distribution with large
variance. As expected, it fails to discriminate against two narrow
normal \pdf's because the implicit assumption of overall normality for
the alternative hypothesis does not hold in this case. The
distance-to-nearest-neighbor method performs better than the \kolm\ approach
for every test. This confirms our intuitive assumption about enhanced
sensitivity of the distance-to-nearest-neighbor method to deviations of
data from an expected \pdf. We note that the proposed
distance-to-nearest-neighbor method is versatile as it provides some
level of discrimination against every alternative hypothesis, although
by no means should it be expected to provide the best discrimination against
every alternative hypothesis.

\simplex{htbp}{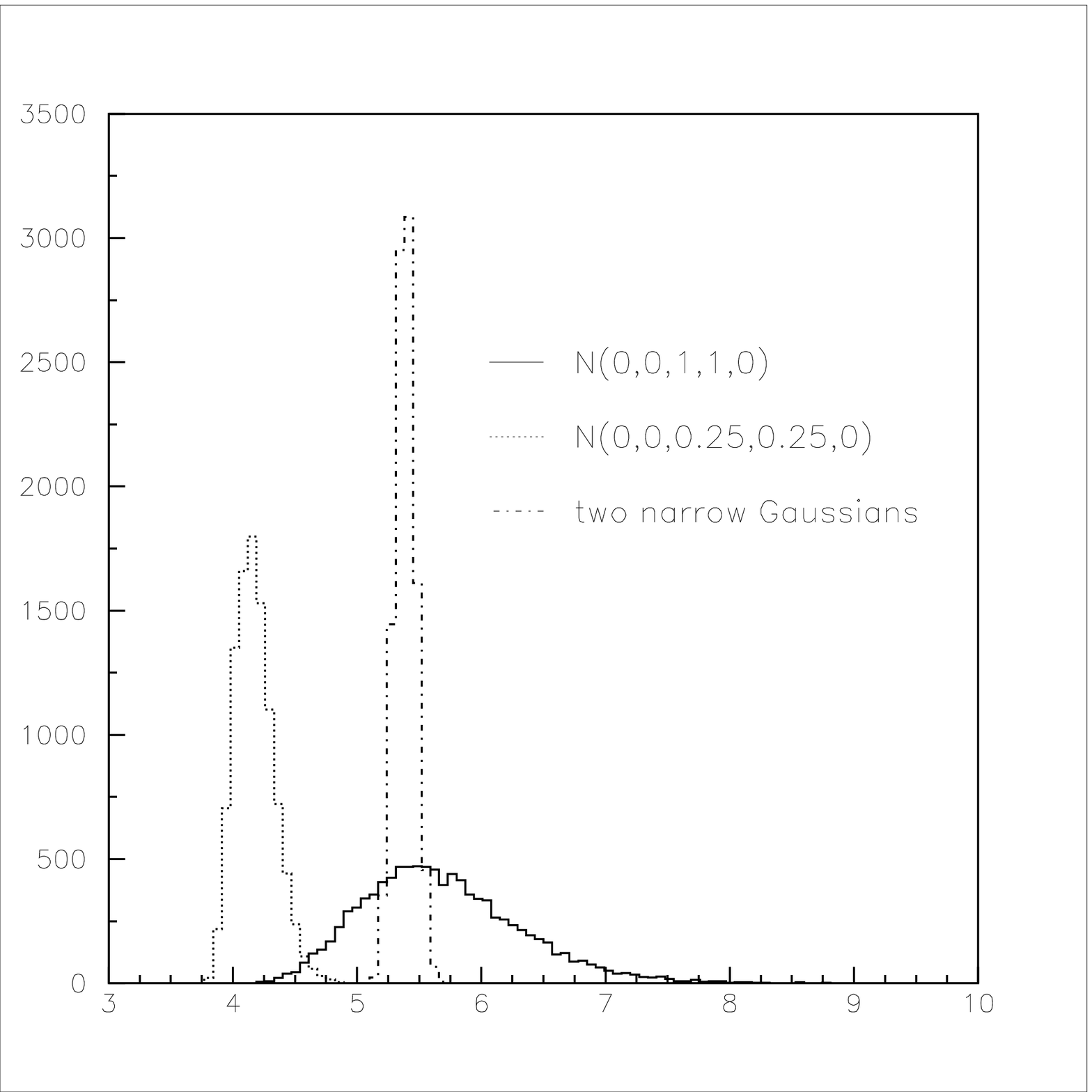}{4lkh}{
$-2\log\lkh_0$ under the null hypothesis $N(0,0,1,1,0)$ for the four
\pdf's discussed in the text. A histogram for the uniform \pdf\ is
not shown because it is far to the right. }

\simplex{htbp}{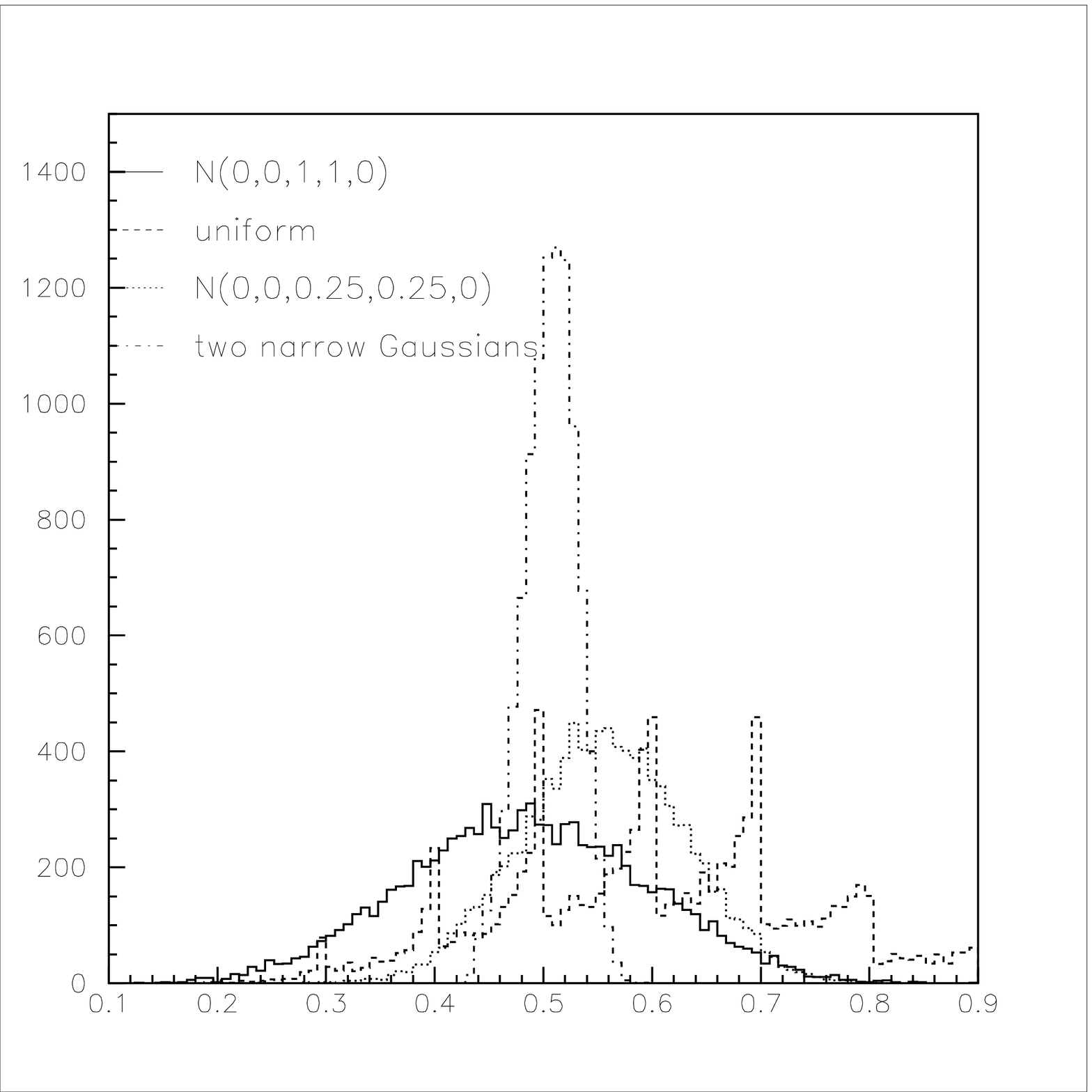}{4kolm}{
The \kolm\ statistic under the null hypothesis $N(0,0,1,1,0)$ for the four
\pdf's discussed in the text. }

\simplex{htbp}{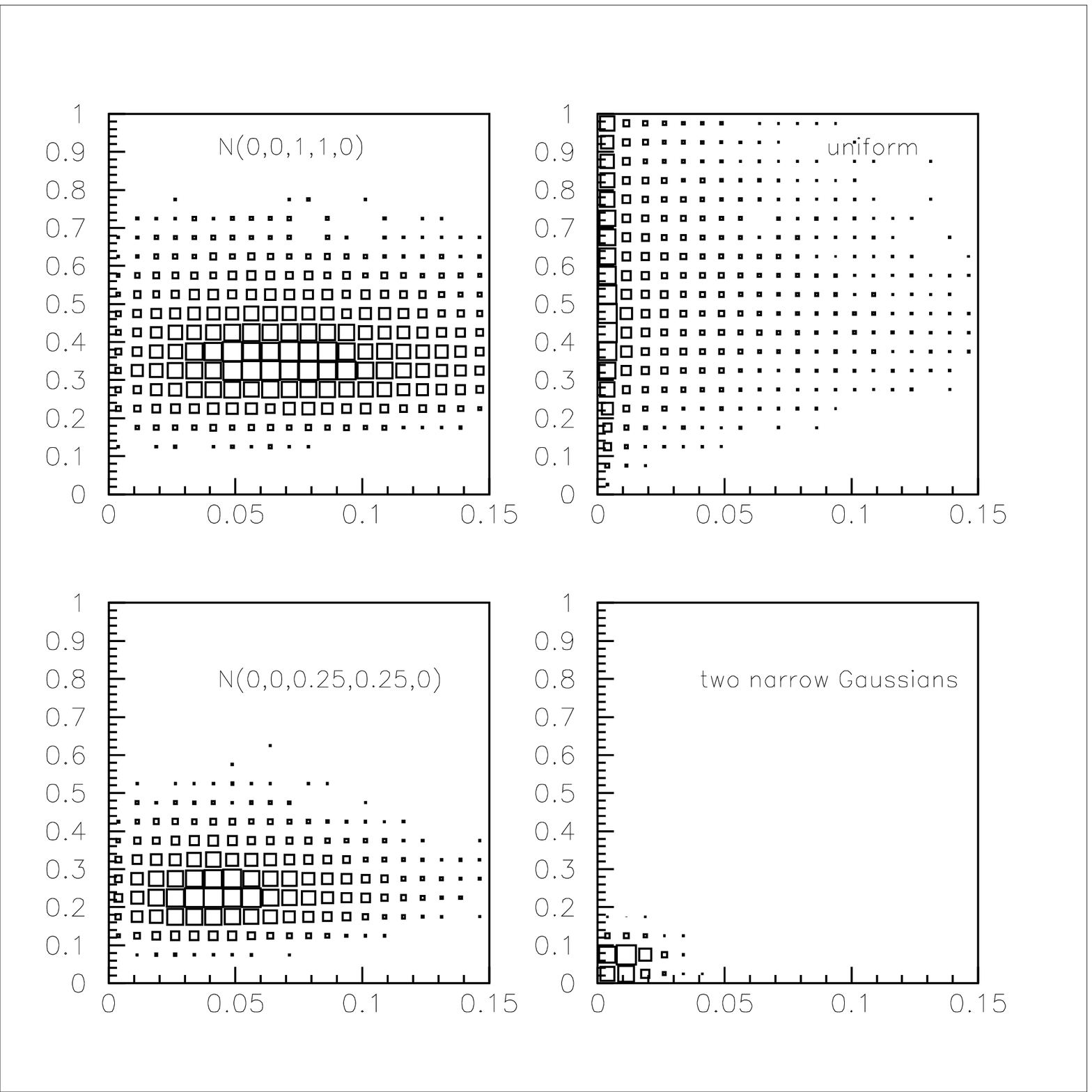}{4dist}{
Maximal vs minimal distance to nearest neighbor computed
under the null hypothesis $N(0,0,1,1,0)$ for the four
\pdf's discussed in the text. The histogram for the uniform \pdf\
shows a very narrow peak at the left edge of the plot. }

\simplex{htbp}{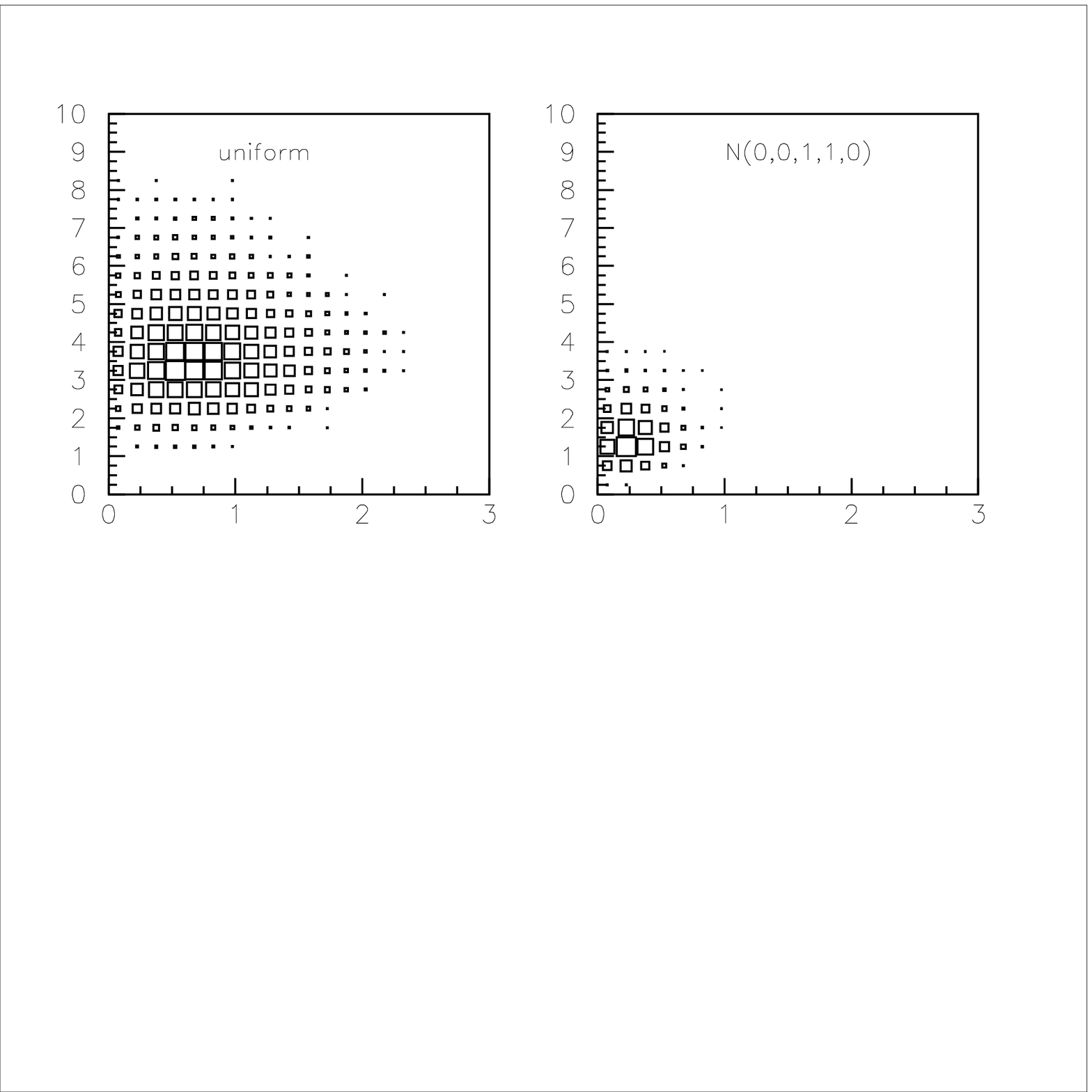}{udist}{
Maximal vs minimal distance to nearest neighbor computed
under the uniform null hypothesis for the uniform \pdf\
defined on a square 
$-5\leq x\leq 5;\ -5\leq y\leq 5$ and $N(0,0,1,1,0)$ . }

\begin{table}[bthp]
\begin{center}
\caption{Confidence levels (CL) and Type~II errors for the maximum
likelihood value (MLV), \kolm\ (KS), and distance-to-nearest-neighbor
(DTNN) tests. DTNN Type~II errors can be reduced for the
$N(0,0,1,1,0)$-vs-$N(0,0,0.25,0.25,0)$ test by imposing a
two-dimensional linear cut on the distributions shown in
Fig.~\ref{fig:4dist}. Such a cut was not used here because these
values are for illustration only. }
\small
\begin{tabular}{|c|c|c|c|c|c|}\hline
Test            & CL                   
& \multicolumn{3}{c|}{Type~II error}    & Comment\\ \cline{3-5}
& & MLV test & DTNN test & KS test & \\ \hline

$N(0,0,1,1,0)$ 
vs uniform & 95\% & 0.0\% & 27.2\%   & 66.2\%  & cutting on minimal \\
           & 50\% & 0.0\% & 0.5\%    & 20.4\%  & distance for DTNN \\ \hline

$N(0,0,1,1,0)$ 
vs $N(0,0,0.25,0.25,0)$ 
& 95\% & 0.8\%  & 55.9\% & 93.7\% & cutting on maximal  \\
& 50\% & 0.0\%  & 6.7\%  & 17.1\% &  distance for DTNN  \\ \hline

$N(0,0,1,1,0)$ 
vs two narrow normal \pdf's 
& 95\% & 100\%  & 0.0\%    & 100\%  & cutting on maximal \\
& 50\% & 97.1\% & 0.0\%    & 18.4\% & distance for DTNN \\ \hline

uniform vs $N(0,0,1,1,0)$ 
& 95\% & N/A & 0.6\% & 75.6\% & cutting on maximal \\
& 50\% &     & 0.0\% & 0.1\%  & distance for DTNN \\ \hline
\end{tabular}
\end{center}
\end{table}
\normalsize

\section{Example: Evidence for $B\to K^{(*)}l^+l^-$ at \babar}

We apply the proposed distance-to-nearest-neighbor method to results of a
$B\to K^{(*)}l^+l^-$ study~\cite{kll} at \babar. In this study, eight
$B\to K^{(*)}l^+l^-$ decays were investigated. Signal rate and upper
limit estimates were obtained for these eight decays. We concentrate
on two modes with measured signal yields: $N(B^+\to K^+e^+e^-) =
14.4_{-4.2}^{+5.0}$ and $N(B^+\to K^+\mu^+\mu^-) = 0.5_{-1.3}^{+2.3}$
(statistical errors only). The former can be described as a
``significant measurement'' while the latter can be used to set an
upper limit.

Signal yields in this analysis are obtained using unbinned maximum
likelihood fits to two-dimensional distributions of energy versus mass
shown in Fig.~\ref{fig:kll2d}. The two-dimensional background is
modeled by the \pdf
\begin{equation}
\label{eq:kll_background}
f(\Delta E,m_{ES}) = A \cdot \exp(s\Delta E) \cdot m_{ES} 
                     \sqrt{1-\frac{m_{ES}^2}{E_b^2}}
    \cdot \exp\left[-\xi\left( 1-\frac{m_{ES}^2}{E_b^2} \right)\right]\ ,
\end{equation}
where $\Delta E=E_{Kll}-E_b$ is the difference between the energy of
the $B$ candidate and the beam energy, $E_b=5.29$~GeV$/c^2$; $m_{ES}$
is the beam-constrained mass of the $B$ candidate; $s$ and $\xi$ are
shape parameters; and $A$ is a factor needed for proper
normalization of the \pdf. The signal shape is modeled by a
normal-like function whose specific analytic expression is not
important for this exercise.

\duplex{htbp}{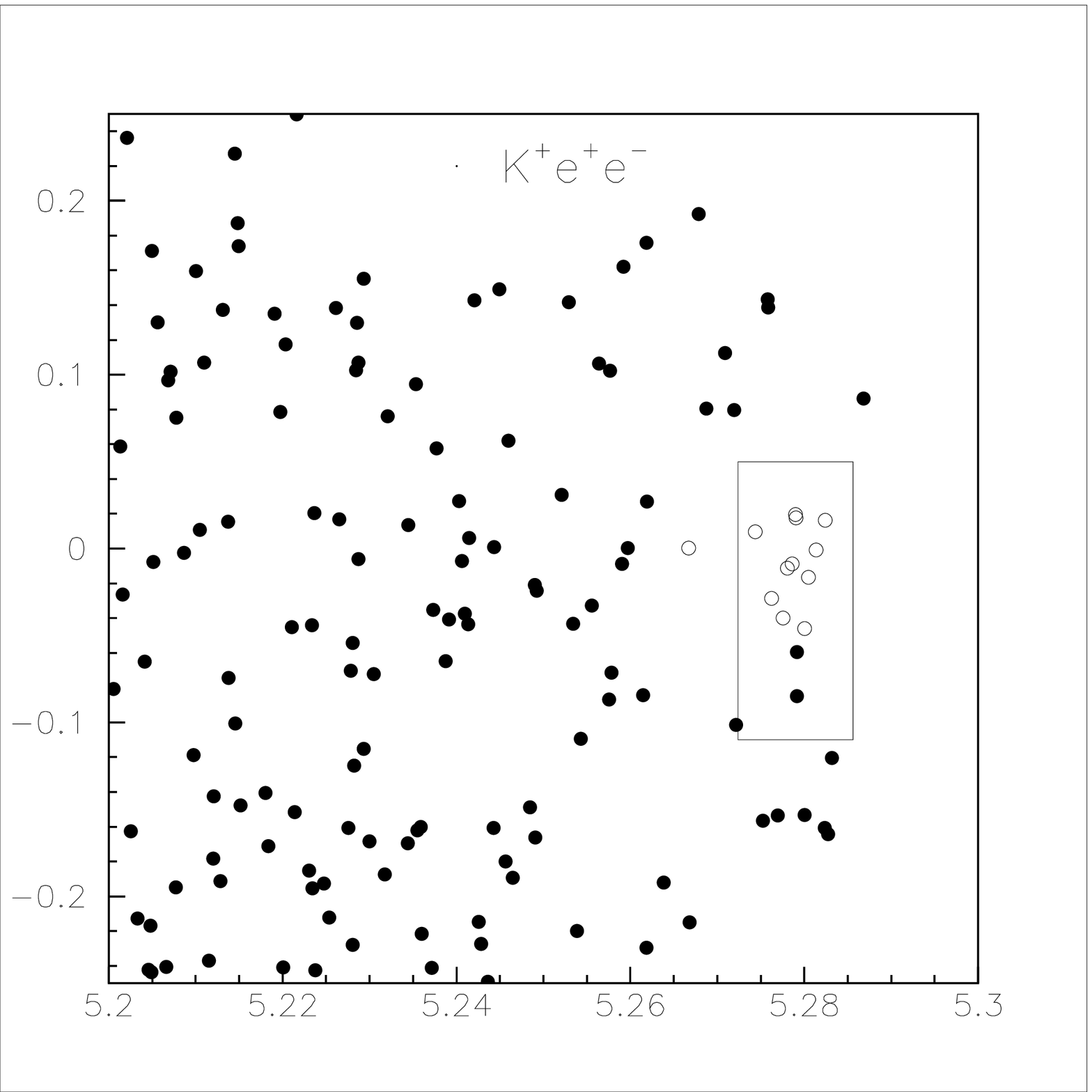}{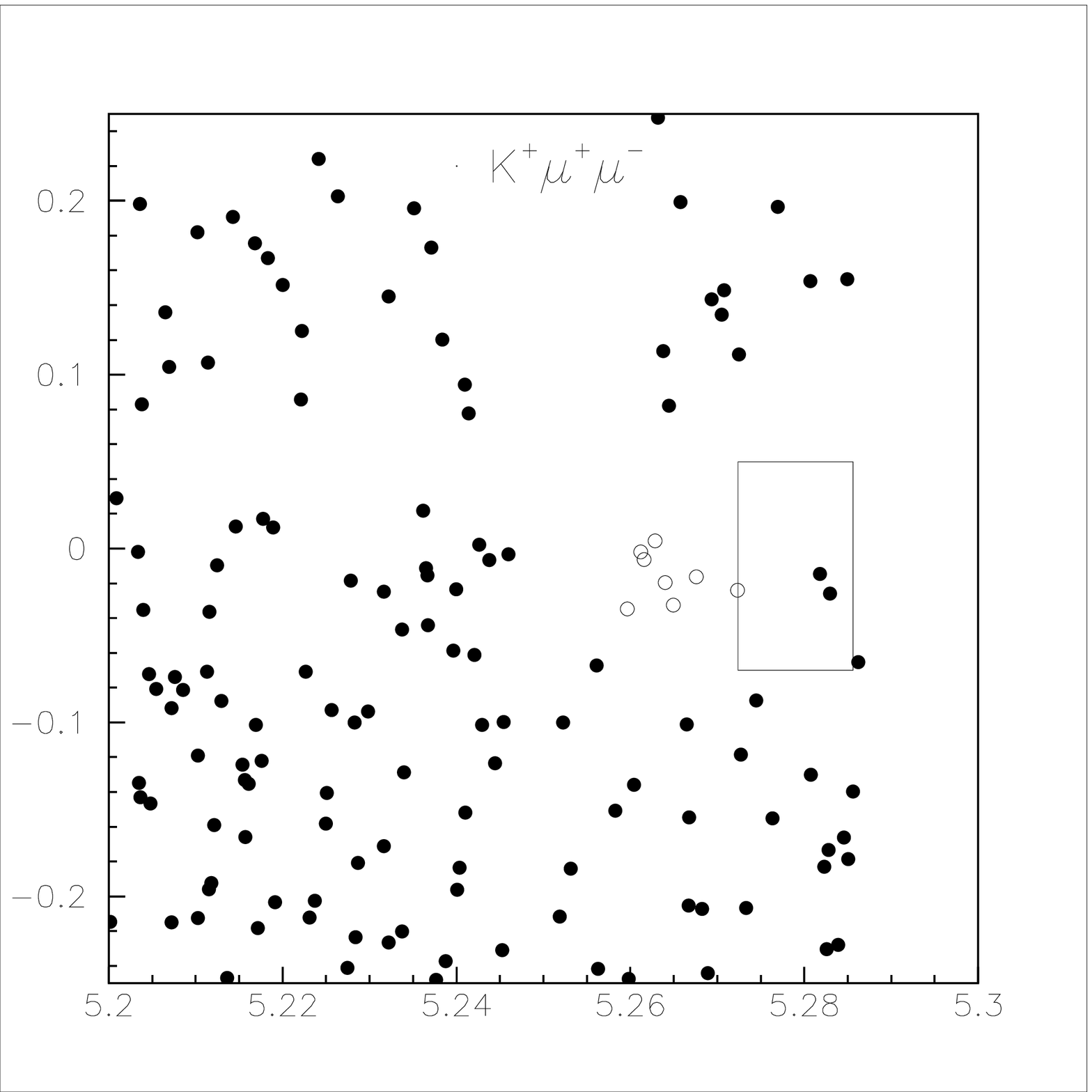}{kll2d}{ Difference $\Delta E$
(GeV) between the energy of the reconstructed $B$ candidate and the
beam energy versus beam-constrained mass $m_{ES}$ (GeV$/c^2$) of the
reconstructed $B$ candidate. Data for the $B^+\to K^+e^+e^-$ decay are
shown on the left, and data for the $B^+\to K^+\mu^+\mu^-$ decay are
shown on the right. Signal regions are shown with boxes. Data clusters
that give maximal deviations from the expected \pdf's are shown with
open circles. }

We ask the following question: How consistent are the observed data
with the background \pdf? In other words, we compute goodness-of-fit
values assuming that all events come from the background. The
background \pdf~(\ref{eq:kll_background}) is smooth while a hypothetical
signal is expected to manifest itself through accumulation of events
in a small region of the two-dimensional plot. In this case, the
alternative hypothesis can be reasonably stated as ``presence of peaks
in the data''. Presence of peaks in the data would result in a smaller
minimal distance $d_i^{(m)}$ to $m$ nearest neighbors than the one
expected from the smooth background \pdf.

To estimate the goodness-of-fit, we transform the background
\pdf~(\ref{eq:kll_background}) to uniform using Eq.~(\ref{eq:tounif}),
generate 10,000 MC experiments and determine the goodness-of-fit as a
fraction of these MC experiments where the minimal distance
$d_i^{(1)}$ to nearest neighbor is less than the one observed in the
data. We conclude that the $B^+\to K^+e^+e^-$ and $B^+\to
K^+\mu^+\mu^-$ data are consistent with the fit at the 51\% and 80\%
level, respectively. At this point, there is no indication of any
peaks in the data.

Now we repeat the exercise
described in the previous paragraph for $d_i^{(m)}$,
$m>1$. Goodness-of-fit values are plotted versus $m$ for both $Kll$
modes in Fig.~\ref{fig:kllcl}. For a cluster of size 12, we estimate
that the $B^+\to K^+e^+e^-$ data are consistent with the fit only at
the 0.13\% level. At the same time, the goodness-of-fit for the
$B^+\to K^+\mu^+\mu^-$ data does not depend dramatically on the
cluster size. The lowest goodness-of-fit value of 6.8\% for the
$B^+\to K^+\mu^+\mu^-$ data corresponds to the test with clusters of
size 8.

We conclude that the $B^+\to K^+e^+e^-$ data are inconsistent with the
background density. Not surprisingly, the data cluster that gives the
maximal deviation from the background \pdf\ consists mostly of points
located inside the signal region.

\duplex{htbp}{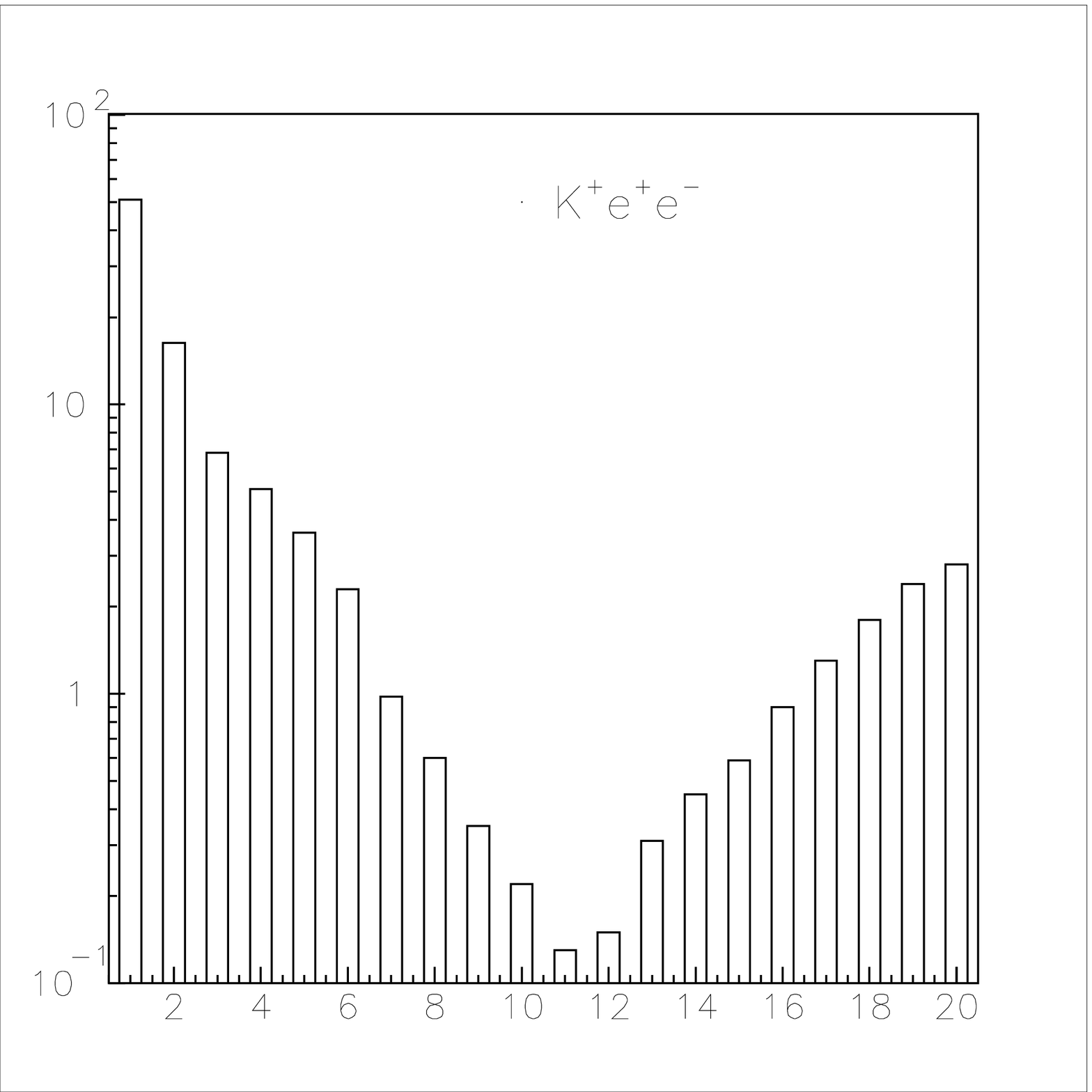}{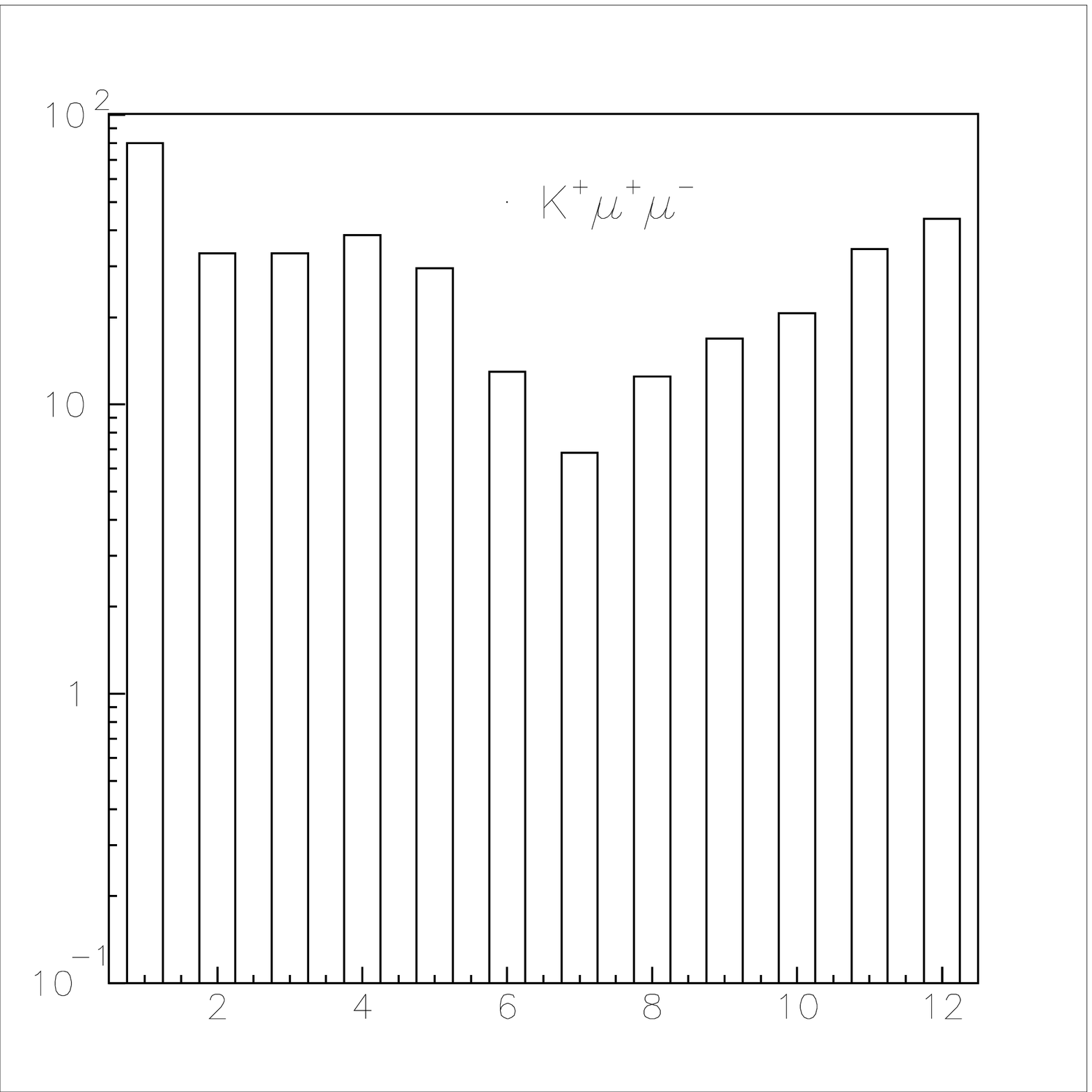}{kllcl}{ Goodness-of-fit (\%)
versus number of nearest neighbors (cluster size minus one) included in
the goodness-of-fit calculation for the $B^+\to K^+e^+e^-$ data (left)
and $B^+\to K^+\mu^+\mu^-$ data (right). }

\section{Summary}

We have proposed a new method for estimation of goodness-of-fit in
multidimensional analysis using a distance-to-nearest-neighbor test of
uniformity. This procedure is recommended as a more versatile tool
than the maximum likelihood methods for a vague generic alternative
hypothesis. However, if the alternative hypothesis is stated in more
specific terms, other methods may be superior.

\begin{acknowledgments}
Thanks to Frank Porter for reviewing this note. Thanks to Art Snyder
and Mike Sokoloff for comments. Thanks to Anders Ryd for providing
details of the $B\to Kll$ analysis.
\end{acknowledgments}

\end{document}